\documentclass{aa}
\usepackage{graphicx}
\usepackage{txfonts}
\usepackage{natbib}
\usepackage{CJK}
\usepackage{graphicx}
\bibpunct{(}{)}{;}{a}{}{,}

\usepackage{lscape}
\usepackage{rotating}

\usepackage[colorlinks,linkcolor=blue,anchorcolor=blue,citecolor=blue,]{hyperref}
\usepackage[switch]{lineno}

\usepackage{caption}
\begin{document}
\begin{CJK*}{UTF8}{gbsn}
\title{Possible origins of the two $\gamma$-ray components C0 and C1 in IC 443} 
\author{Yujun Xian\inst{1},  
Huan Yu\inst{2}, 
Bo Wang\inst{3},
Jun Fang\inst{1}
}
\offprints{\email{fangjun@ynu.edu.cn; wangbo@ynao.ac.cn; yuhuan.0723@163.com}}

\institute{International Centre of Supernovae (ICESUN), Yunnan Key Laboratory of Supernova Research, School of Physics and Astronomy,Yunnan University, Kunming 650091, China  
\and Department of Physical Science and Technology, Kunming University, Kunming 650214, China   
\and International Center of Supernovae (ICESUN), Yunnan Key Laboratory of Supernova Research, Yunnan Observatories, Chinese Academy of Sciences, Kunming 650216, China
}

\date{Received; accepted}
\abstract 
{Recent LHAASO observations of IC~443 reveal that its high-energy $\gamma$-ray emission can be decomposed into two components with clearly distinct spatial morphologies and spectral properties, namely the compact component C0  and component C1 with an angular extent of 0.67$^\circ$.}
{We aim to explain the $\gamma$-ray emission mechanisms responsible for the two components in IC~443 and to further investigate their respective possible astrophysical counterparts.}
{We model  components C0 and C1 independently using two different physical scenarios.  For component C0, we adopt an SNR-MC hadronic model that accounts for finite source-size effects to describe the escape and diffusion of high-energy particles and their $\gamma$-ray production via pp interactions with the surrounding molecular clouds. For component C1 , we employ a spatially dependent multi-zone leptonic transport model and describe the evolution of high-energy electrons under the combined action of advection, diffusion, and energy losses using a Fokker--Planck-type equation.}
{The model reproduces the SEDs of components C0 and C1  well.  Component C0 exhibits a pronounced $\pi^0$-decay feature, and the maximum energy of its escaping particles reaches the PeV scale. The $\gamma$-ray emission from component C1  can be explained by inverse Compton scattering. Its diffusion coefficient is approximately two orders of magnitude lower than that in the surrounding interstellar medium, and no significant synchrotron counterpart is detected.}
{The $\gamma$-ray emission from IC~443 involves the combined action of multiple physical mechanisms. Component C0 may be a potential PeVatron candidate, whereas component  C1  may correspond to a potential TeV halo candidate.}

\keywords {ISM: supernova remnants -- gamma rays: ISM -- diffusion -- radiation mechanisms: non-thermal }

\titlerunning{C0 and C1 $\gamma$-ray components in IC~443}
\authorrunning{Xian et al.}

\maketitle

\section{Introduction}
\label{sect:1}
Massive stars undergo violent explosions in the late stages of their evolution, ejecting large amounts of material at high velocities into the surrounding interstellar medium. Through interactions with the ambient medium,  these highly energetic ejecta gradually evolve into supernova remnants (SNRs) (\citealp{Reynolds2008}; \citealp{Vink2012}). As key objects in high-energy astrophysics, SNRs are widely regarded as one of the principal sources of Galactic cosmic rays (CRs)  and as potential PeVatrons capable of accelerating CRs up to PeV energies  (\citealp{Mitchell2021}; \citealp{Fang2022}).

A peculiar SNR, IC~443 (G189.1+3.0), is regarded as one of the brightest $\gamma$-ray sources with a prominent extended structure in the Galaxy.  Located at a distance of about 1.5~kpc \citep{Fesen1984}, and with an estimated age of 3$-$30~kyr, it is generally classified as a middle-aged SNR (e.g. \citealp{Petre1998}; \citealp{Olbert2001}; \citealp{Lee2008}). Observational and theoretical studies suggest that IC 443 most likely originated from a core-collapse supernova explosion and exhibits the characteristic properties of a mixed-morphology SNR, including one of the best-studied SNR-MC interaction regions (e.g. \citealp{Rho1998}; \citealp{Su2014}; \citealp{Fuda2023}). Previous studies have also suggested a possible physical association between IC 443 and the pulsar wind nebula (PWN) CXOU J061705.3+222127 (\citealp{Swartz2015}; \citealp{Ustamujic2021}).

IC 443 is one of the best-studied SNRs at $\gamma$-ray energies. Its $\gamma$-ray emission was first detected by the MAGIC telescope and reported as the point-like source MAGIC J0616+225 \citep{MAGIC2007}.  Subsequent observations by VERITAS showed that the source in fact exhibits spatially extended $\gamma$-ray emission, with its brightest region spatially coincident with the distribution of dense MCs \citep{VERITAS2009}. In addition, Fermi-LAT observations revealed the characteristic emission produced by $\pi^0$ decay, indicating the presence of CR protons accelerated to high energies in the IC 443 region (\citealp{Abdo2010}; \citealp{Fermi2013}).

\begin{figure}[t]
    \centering
    \includegraphics[width=\columnwidth]{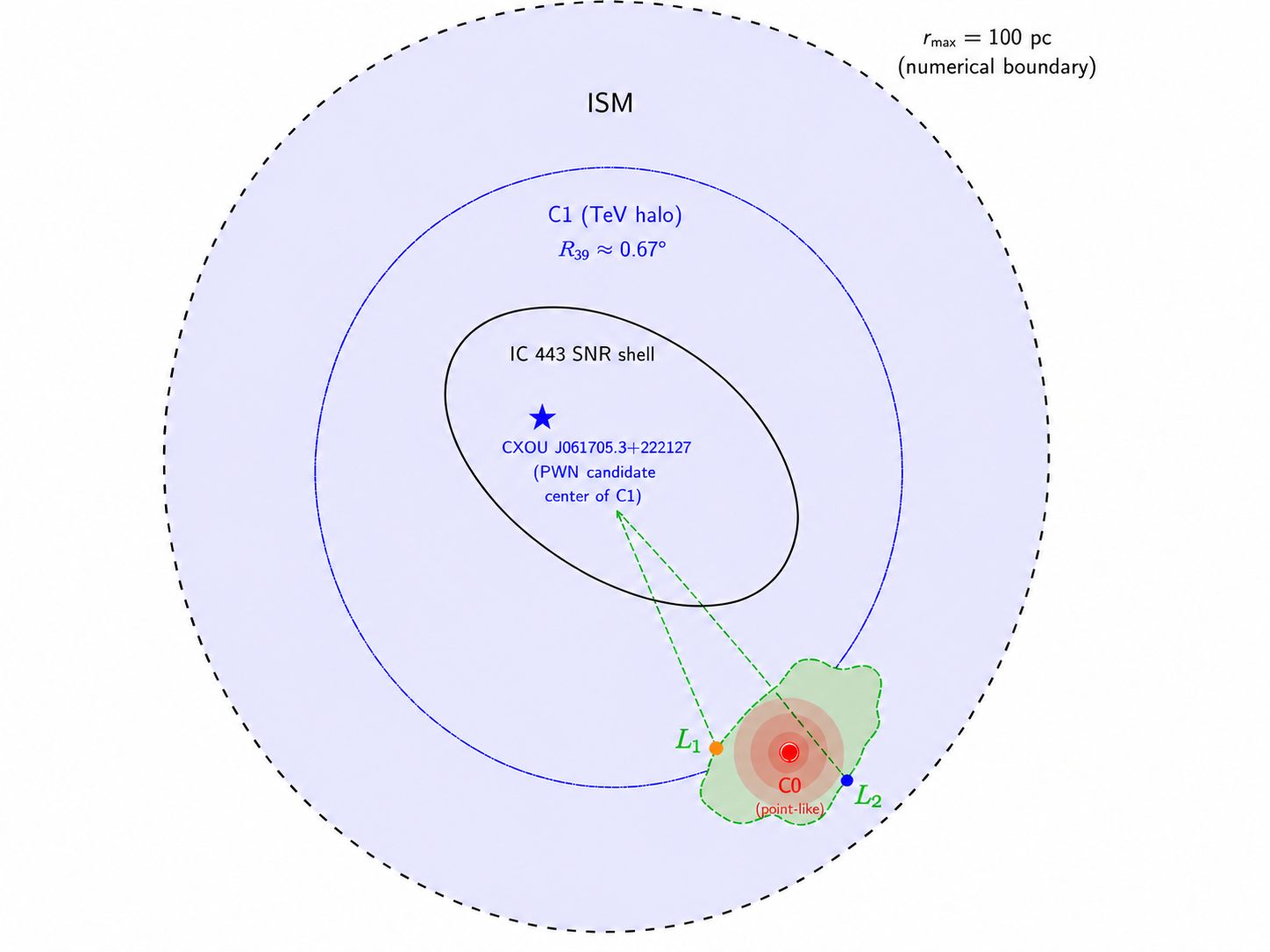}
    \caption{Schematic illustration of the relationship between the two $\gamma$-ray components C0 and C1 in IC 443.}
    \label{figure1}
\end{figure}

Recently, benefiting from its broad energy coverage,  LHAASO performed a more detailed analysis of the high-energy $\gamma$-ray emission from the IC~443 region and further decomposed the previously unresolved overall emission into two components, C0 and C1 \citep{LHAASOCollaboration2025}.  The observations show that  component C0,  located inside the remnant,  has a relatively compact spatial distribution and is statistically consistent with a point source,  whereas component C1  exhibits a significantly more diffuse morphology with an extension scale of about $0.67^\circ$,  substantially larger than that of C0.  \citet{LHAASOCollaboration2025} further found that the SED of C0 connects smoothly with that of the compact source 4FGL~J0617.2+2234e detected by Fermi-LAT \citep{Abdollahi2020}.  In addition,  component C0  exhibits radiative features associated with interactions with molecular clouds and shows a prominent $\pi^0$-decay bump,  indicating that its $\gamma$-ray emission most likely originates from hadronic processes in the SNR-MC interaction system  (\citealp{Tang2011}; \citealp{Li2012}; \citealp{Tang2015}).  In contrast,  C1 is spatially coincident with the extended source 2FGES~J0618.3+2227 discovered by Fermi-LAT \citep{Abdollahi2024}.  Although a certain amount of MC gas is present in the C1 region,   the $\gamma$-ray morphology does not show a clear spatial correspondence with the MC distribution \citep{LHAASOCollaboration2025}. This feature disfavours a hadronic interpretation for the $\gamma$-ray emission of C1 and instead suggests that it is more likely to be of leptonic origin.

To understand the $\gamma$-ray emission mechanism of IC~443, a variety of theoretical frameworks have been proposed in previous studies (e.g. \citealp{Baring1999}; \citealp{Bartko2008}; \citealp{Uchiyama2010}; \citealp{Tang2014}; \citealp{Tang2019}). These works explored the origin of the high-energy emission from this source from different perspectives and modelled  the possible particle acceleration and radiative processes responsible for it.  However,  systematic theoretical  modelling of the latest LHAASO observations remains relatively limited.    
To interpret the two emission components C0 and C1, identified by LHAASO in the IC 443 region, we model them separately as two independent $\gamma$-ray sources. For the C0 component, we adopt an SNR--MC hadronic model to investigate the $\gamma$-ray emission produced by interactions between high-energy protons escaping from the SNR and the surrounding molecular clouds. For the C1 component, we independently employ a spatially dependent multi-zone leptonic model to investigate the transport of high-energy electrons and the associated inverse Compton (IC) emission. We assume that there is no coupling between the two models in terms of particle transport or radiative processes. The geometrical configurations and characteristic spatial scales associated with these two independent models are illustrated schematically in Figure~\ref{figure1}.

The SNR-MC hadronic model simultaneously accounts for the evolution of the SNR and the escape of high-energy protons.  After escaping into the surrounding MCs,  these protons can generate $\gamma$-ray emission through pp interactions  \citep{Yang2015}. In this work,  we investigate the $\gamma$-ray emission of component C0  in IC~443 based on the hadronic emission model developed by \citet{Xian2025RAA}. It should be noted that,  in the morphological analysis of component C0,  \citet{LHAASOCollaboration2025} treated this component as a point source limited by the instrumental point spread function in order to improve the statistical significance and the stability of the spectral fitting.  However,  under realistic astrophysical conditions, the interaction region between the supernova-remnant shock and the surrounding MCs is unlikely to be truly point-like,  and is instead expected to exhibit a shell-like morphology or a certain degree of spatial extension. Therefore, the point-source approximation may not adequately capture its intrinsic spatial distribution  \citep{Telezhinsky2012}.

Motivated by this consideration, we incorporate the effect of the finite source size of the SNR into the SNR-MC hadronic model (\citealp{Li2010}; \citealp{Ohira2011}),  and treat  component C0 of IC~443 as a finite-sized volume source.  This treatment allows for a more realistic description of the spatial and spectral distributions of CR injection, diffusion, and $\gamma$-ray emission. Within this framework,  component C0  of IC~443 can be interpreted as hadronic emission associated with a potential Galactic PeVatron.

The multi-zone leptonic model is spatially dependent and usually assumes a spherically symmetric geometry (\citealp{Lu2017ApJ}; \citealp{Lu2019AA}).  Although component C1  partially overlaps spatially with the known SNR G189.6+03.3, no significant radiation has been detected towards this remnant \citep{Alfaro2025}, and there is currently no clear observational evidence supporting a direct physical connection between the C1 emission and this remnant. To reduce the number of free model parameters and thereby lower the associated uncertainties,  we adopt the multi-zone leptonic framework used by \citet{Xian2026ApJ} and assume that the radiation in this region originates entirely from the pulsar wind nebula powered by CXOU~J061705.3+222127. Within this framework, we investigate the radiative properties of component C1  in IC~443.

Within this multi-zone leptonic framework, high-energy electrons are injected near the pulsar termination shock, and their subsequent evolution is described by a Fokker--Planck-type transport equation \citep{van2018arXiv}. The evolution of these electrons takes into account physical processes such as convection, diffusion, and energy losses, thereby leading to a spatially varying distribution of the electron number density.  Based on the analysis of \citet{LHAASOCollaboration2025},  we suggest that  component C1  of IC~443 is likely a TeV halo candidate. 

The structure of this paper is as follows. In Section~\ref{sect:2}, we introduce the SNR--MC hadronic model and the spatially dependent multi-zone leptonic model adopted in this work.  In Section~\ref{sect:3}, we present the fitting results for components C0 and C1  of IC~443 based on these models. In Section~\ref{sect:4}, we summarise the main conclusions and provide further discussion.

\section{Model Description}
\label{sect:2}
\subsection{SNR-MC hadronic model}
\label{sect:2.1}
In this section, we briefly review the SNR--MC hadronic model adopted by \citet{Xian2025RAA}. A detailed description of the model construction can be found in Section~2 of \citet{Xian2025RAA}. At the shock front of an SNR, particles can gain energy through diffusive shock acceleration and gradually escape from the acceleration region into the surrounding medium \citep[e.g.,][]{Bell1978,Malkov2005,Yang2022,MAGIC2023}. Within this framework, the SNR is treated as a spherical system with a finite spatial extent, and the spectrum of particles escaping from the remnant is given by
\begin{equation}
\label{eq1}
N_{\text{esc}}(p) =
\begin{cases} 
 N_0\,p_{\text{br,1}}^{-\frac{\beta}{\alpha}} p^{-s} & \quad  p \leq p_{\text{br,1}} ,\\[1mm]
 N_0\,p^{-(s+\frac{\beta}{\alpha})}\exp\left(-\frac{p}{{p_{\text{cut}}}}\right) & \quad p > p_{\text{br,1}},
\end{cases}
\end{equation}
where $N_{\mathrm{esc}}(p)$ denotes the escape spectrum of CR particles at momentum $p$,  $N_0$ is a constant normalization factor,  and $s$ is the spectral index,  with $\beta = 3(3 - s)/2$. $p_{\mathrm{cut}}$ is the cutoff momentum, and $p_{\mathrm{br},1}$ denotes the break momentum. 
The parameter $\alpha$ is the index that describes the evolution of the cosmic-ray escape momentum, namely the evolution of the characteristic escape momentum of cosmic rays during the expansion of the SNR. Following the modelling of IC 443 by \citet{Ohira2011} , we adopt $\alpha = 6.5$.

CR particles escaping from the SNR undergo pp interactions with the surrounding MCs.  In this work, the MC is treated as a dense gas structure of finite volume occupying the region $L_1 \leq r \leq L_2$.  After integration over the MC volume, the resulting CR energy spectrum is given by the following equation  \citep{Xian2025RAA}
\begin{equation}
\label{eq2}
F(p) = \eta\int_{L_1}^{L_2} f_{\text{}}(t_{\text{age}}, {{r}} , p) 4 \pi{{r}^2} \, {{dr}},
\end{equation}
where $f(t_{\rm age}, r, p)$ denotes the spatial distribution function of escaped CR particles with momentum $p$ at a radial distance $r$ from the centre of the SNR at the current age $t_{\rm age}$.  $\eta$ represents the fraction of escaped particles that collide with the MC.  $\eta N_0$ can be constrained by the observed radiation flux. The final analytical expression for the CR energy spectrum is given by \citep{Ohira2011}
\begin{equation}
\label{eq3}
\begin{aligned}
F(p) &= \frac{N_{\rm esc}(p)}{2}\frac{R_d(t,p)}{\sqrt{\pi}R_{\rm esc}(p)}
\Bigg[
\exp\!\left(-\left(\frac{L_1-R_{\rm esc}(p)}{R_d(t,p)}\right)^2\right) \\
&\quad - \exp\!\left(-\left(\frac{L_2-R_{\rm esc}(p)}{R_d(t,p)}\right)^2\right) \\
&\quad - \exp\!\left(-\left(\frac{L_1+R_{\rm esc}(p)}{R_d(t,p)}\right)^2\right) \\
&\quad + \exp\!\left(-\left(\frac{L_2+R_{\rm esc}(p)}{R_d(t,p)}\right)^2\right) \\
&\quad + \mathrm{erf}\!\left(\frac{L_2-R_{\rm esc}(p)}{R_d(t,p)}\right)
- \mathrm{erf}\!\left(\frac{L_1-R_{\rm esc}(p)}{R_d(t,p)}\right) \\
&\quad + \mathrm{erf}\!\left(\frac{L_2+R_{\rm esc}(p)}{R_d(t,p)}\right)
- \mathrm{erf}\!\left(\frac{L_1+R_{\rm esc}(p)}{R_d(t,p)}\right)
\Bigg],
\end{aligned}
\end{equation}
where \( \operatorname{erf}(x) = \frac{2}{\sqrt{\pi}} \int_0^x e^{-y^2} \, \mathrm{d}y \) is the error function. $R_{\mathrm{esc}}$ denotes the escape radius of the particles,  and the diffusion radius $R_d$ is given by \citep{Aharonian1996}
\begin{equation}
\label{eq4}
R_d(t, p) = \sqrt{4 D_{\text{}}(p) \left[ t - t_{\text{esc}}(p) \right]},
\end{equation} 
where $t$ is the time variable, $t_{\mathrm{esc}}(p)$ represents the escape time of the particles, and the diffusion coefficient $D(p)$ is given by \citep{Li2010}
\begin{equation}
\label{eq5}
D(p) = 10^{28} \chi \left( \frac{cp}{10 \, \text{GeV}} \right)^\delta \, \text{cm}^2 \, \text{s}^{-1},
\end{equation}
where $\chi$ is the normalization factor, $c$ is the speed of light, and $\delta$ is the index that describes the energy dependence of the diffusion coefficient.

\subsection{Multi-zone leptonic model}
\label{sect:2.2}

\begin{figure}[t]
    \centering
    \includegraphics[width=0.85\columnwidth]{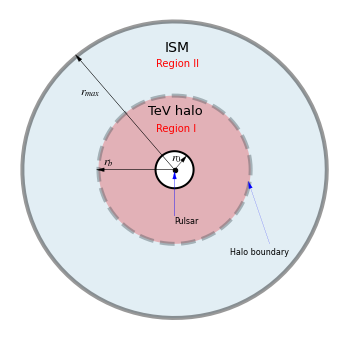}
    \caption{Schematic illustration of the multi-zone leptonic model.}
    \label{figure2}
\end{figure}

To investigate the $\gamma$-ray emission from component C1  in IC 443,  we adopt a multi-zone (multishell) leptonic model.  A detailed description of the model construction can be found in Section~2 of \citet{Xian2026ApJ}.  The geometrical configuration of the model is illustrated in figure~\ref{figure2}, with the pulsar located at the centre of the system. The model consists of two main regions characterized by different diffusion coefficients, namely the TeV halo region close to the pulsar, denoted as Region~I, and the outer interstellar medium region, denoted as Region~II. The white region in the figure generally corresponds to the near-source injection zone, whose physical properties are primarily associated with particle injection rather than the main processes of particle transport and evolution. Owing to the complexity of the physical environment in this region, directly including it in the model would introduce a large number of poorly constrained parameters. This region is not modelled in this work, consistent with the approach of \cite{Kundu2024}.  
The radius $r_0$ therefore represents the fixed numerical inner boundary of the resolved transport region rather than the time-dependent physical termination-shock radius.
Physically, $r_0$ represents an effective particle injection boundary located near or immediately downstream of the termination shock, rather than the physical termination-shock radius itself. The actual termination-shock radius $r_{\rm ts}(t)$ is expected to evolve with the PWN dynamics. However, the compact region between $r_{\rm ts}(t)$ and $r_0$, including particle acceleration, pulsar-wind thermalisation, and magnetic-field evolution, is not spatially resolved in the present phenomenological model. Its effects are effectively incorporated into the injected particle spectrum $Q(E_e,t)$ at the innermost numerical shell, and the transport calculation starts from the resolved region beyond this effective injection boundary. To reduce the number of poorly constrained model parameters and to focus on the large-scale particle transport and radiative properties of the extended C1 component, the present model does not explicitly follow the dynamical evolution of the termination shock.

In the spherically symmetric system,  the evolution of the electron distribution $N_{\rm e}(r, E_{\rm e}, t)$ is described by the following  Fokker-Planck-type particle transport equation \citep{van2018arXiv,Kundu2024}
\begin{align}
\label{eq6}
\frac{\partial N_e}{\partial t} = & - \mathbf{V} \cdot (\nabla N_e) + \kappa \nabla^2 N_e \nonumber \\
& + \frac{1}{3} (\nabla \cdot \mathbf{V}) \left( \left[ \frac{\partial N_e}{\partial \ln E_e} \right] - 2 N_e \right) \nonumber \\
& + \frac{\partial}{\partial E} \left( \dot{E}_{e, \mathrm{rad}} N_e \right) + Q_{\mathrm{inj}},
\end{align}
where  $V$ denotes the bulk velocity of the particles,  and $\kappa(r, E_{\mathrm{e}})$ is the diffusion coefficient that depends on spatial position and particle energy.  $\dot{E}_{\mathrm{e},\mathrm{rad}}$ denotes the total energy loss rate of electrons due to radiative processes, mainly including synchrotron and IC losses. $Q_{\mathrm{inj}}$ describes the particle injection spectrum.  $r$, $E_{\mathrm{e}}$, and $t$ correspond to the radial coordinate,  particle energy,  and time.

In our model, the particle energy losses mainly arise from radiative and adiabatic processes. The radiative energy losses include synchrotron radiation and IC scattering and can be described as follows \citep{Moderski2005}
\begin{align}
\label{eq7}
\dot{E}_{\mathrm{e,rad}}
&=\frac{4}{3}\sigma_{\rm T}c
\left(\frac{E_e}{m_e c^2}\right)^2
\left[U_B+\left(U_{\rm CMB}+U_{\rm NIR}+U_{\rm FIR}\right)\right],
\end{align}
where $\sigma_{\rm T}$ denotes the Thomson scattering cross-section, and $m_e$ is the electron mass. $U = B^2/(8\pi)$ represents the magnetic-field energy density. $U_{\rm CMB}$, $U_{\rm NIR}$, and $U_{\rm FIR}$ denote the energy densities of the cosmic microwave background (CMB), the near-infrared (NIR) photon field, and the far-infrared (FIR) photon field, respectively.

The adiabatic energy-loss rate of the particles can be written as follows \citep{Zhang2008}
\begin{equation}
\label{eq8}
\dot{E}_{e, \mathrm{ad}} = \frac{1}{3} (\nabla \cdot \mathbf{V}) E_e,
\end{equation}
where $\nabla$ denotes the gradient operator.

The pulsar continuously injects energy and high-energy particles into the system through its rotational energy losses,  and the corresponding spin-down luminosity is given by the following equation (\citealp{Fang2010}; \citealp{Martin2012})
\begin{equation}
\label{eq9}
L(t) = L_0 \left( 1 + \frac{t}{\tau_0} \right)^{\frac{n+1}{n-1}},
\end{equation}
where $n$ is the braking index,  $L_0$ and $\tau_0$ denote the initial spin-down luminosity and initial spin-down timescale of the pulsar.  From $\tau_0 = \tau_c - t_{\rm age}$,  the following expression is obtained \citep{Gaensler2006}
\begin{equation}
\label{eq10}
\tau_0 =\frac{P}{(n - 1) \dot{P}} - t_{\text{age}},
\end{equation}
where $\tau_c$ is the characteristic age of the pulsar, $P$ and $\dot{P}$ denote the current spin period and its time derivative,  and $t_{\rm age}$ is the age of the PWN.

In this model,  the innermost spherical shell (the first shell) of the system is treated as the particle injection layer,  where particles are assumed to be uniformly distributed.   The particles then propagate outward under the combined effects of advection and diffusion,   gradually entering the outer shells.  Accordingly, the particle injection term can be expressed as follows \citep{van2018MN}
\begin{equation}
\label{eq11}
Q_{\text{inj}} = \frac{Q(E_e, t)}{V^{1}_{\text{shell}}},
\end{equation}
where $V_{\rm shell}^{1}$ is the volume of the innermost spherical shell,  and $Q(E_e,t)$ represents the electron injection spectrum supplied by the pulsar at time $t$. In Equation~(\ref{eq6}), $Q_{\rm inj}$ is the local injection source term per unit volume. Therefore, $Q(E_e,t)$ describes the instantaneous electron supply from the pulsar, whereas $N_e(r,E_e,t)$ denotes the electron distribution that evolves in space, energy, and time after this injection, as obtained by solving Equation~(\ref{eq6}) with this injection term and including the effects of advection, diffusion, and energy losses. Similar treatments of particle injection and transport have also been adopted in the multi-zone model of \citet{van2018MN}.

In this work,  we adopt an exponential-cutoff power-law (ECPL) distribution \citep{Xian2026ApJ}
\begin{equation}
\label{eq12}
Q(E_e, t) = Q_0(t) E_e^{-s} \exp \left( -\frac{E_e}{E_{\text{cut}}} \right),
\end{equation}
where $s$ is the spectral index of the particle injection spectrum,  and $E_{\mathrm{cut}}$ is the cutoff energy. $Q_0(t)$ denotes the time-dependent normalization factor,  whose value is jointly determined by the pulsar spin-down luminosity $L(t)$ and the fraction of spin-down energy converted into accelerated particle energy \citep{Tian2023}.    $Q_0(t)$ is given by
\begin{equation}
\label{eq13}
\epsilon L(t) = \int_{E_{\text{min}}}^{E_{\text{max}}} Q_0(t) E_e^{-s} \exp \left( - \frac{E_e}{E_{\text{cut}}} \right) \ E_e \, dE_e,
\end{equation}
where $\epsilon$ denotes the constant efficiency with which the spin-down luminosity is converted into particle energy,  and $E_{\max}$ and $E_{\min}$ denote the maximum and minimum particle energies.

Given the markedly different environmental conditions in the inner and outer regions, the diffusion properties of particles in Region~I and Region~II are difficult to describe with a single set of parameters.  Therefore,  in this work,  the diffusion coefficient $\kappa(E_{\mathrm{e}}, r)$ is defined as a piecewise function of particle energy and radial position \citep{Liu2019}
\begin{equation}
\label{eq14}
K(E_{\rm e}, r) =
\begin{cases}
D_1 \left( \dfrac{E_{\rm e}}{E_0} \right)^{\delta_1} & \text{if } r_0 \le r \le r_b,\\[4pt]
D_2 \left( \dfrac{E_{\rm e}}{E_0} \right)^{\delta_2} & \text{if } r_b < r \le r_{\max},
\end{cases}
\end{equation}
where  $E_0$ is the reference energy,  and $D_1$ and $D_2$ denote the normalization constants of the diffusion coefficient in Region~I and Region~II.  The parameters $\delta_1$ and $\delta_2$ are the power-law indices describing the energy dependence of the diffusion coefficient in the two regions.  To reduce the number of free parameters in the model, we adopt the same power-law index for both regions, namely $\delta_1 = \delta_2 = \delta$, and use a Bohm-like diffusion approximation, corresponding to $\delta = 1$. As shown in Figure~\ref{figure1},  $r_b$ is the boundary radius between Region~I and Region~II, marking the transition from the TeV halo region to the outer interstellar medium.  The boundary $r=r_{\max}$ is the maximum distance to which the particles can diffuse, and the particle number density at this boundary naturally satisfies $N_{\rm e}=0$. This requires that the value of $r_{\max}$ be sufficiently large. In this work, we adopt $r_{\max}=100$~pc, which is consistent with \citet{Liu2019}. To further reduce the number of free parameters and to remain consistent with previous studies, we fix $r_0 = 0.02\,\mathrm{pc}$ and $E_0 = 1\,\mathrm{TeV}$ in the numerical calculations, following the treatment of \citet{Kundu2024}.

\begin{table}
\caption{Adopted parameter values for component C0 }
\label{tab1}
\centering
\small
\setlength{\tabcolsep}{4pt}
\renewcommand{\arraystretch}{1.02}
\begin{tabular}{p{4cm}cc}
\hline\hline
Input Parameters & Symbol & Value \\
\hline
Age (kyr)                                 & $t_{\rm age}$           & $30.0$\tablefootmark{a} \\
Distance (kpc)                            & $d$                     & $1.5$\tablefootmark{b} \\
Break momentum(GeV/$c$)                   &$p_{\mathrm{br},1}$      & $62.9$\tablefootmark{c} \\
Spectral index for thermal leakage        & $s$                     & $6.5$\tablefootmark{c} \\
Atomic gas density ($\mathrm{cm}^{-3}$)   & $n_{\mathrm{H}}$        & $20.0$\tablefootmark{d} \\
\hline
\multicolumn{3}{l}{Fitted Parameters} \\
\hline
Spectral index                            & $s$                     & 2.03 \\
Diffusion index                           & $\delta$                & 0.4 \\
Normalization coefficient                 & $\chi$                  & 0.08 \\
Escape spectral break momentum
(GeV/$c$)                                 & $p_{\mathrm{cut}}$      & $5.0 \times 10^{5}$ \\
\hline
\end{tabular}
\tablefoot{
\tablefoottext{a}{Adopted from \citet{Olbert2001}.}
\tablefoottext{b}{Adopted from \citet{Woltjer1972}.}
\tablefoottext{c}{Adopted from \citet{Ohira2011}.}
\tablefoottext{d}{Adopted from \citet{Alfaro2025}.}
}
\end{table}

Equation~(\ref{eq6}) can be written in the following discretized form \citep{van2018MN}
\begin{align}
\label{eq15}
(1 - z + \beta)(N_e)_{i,j+1,k} &= 2 Q_{i,j,1} \Delta t \notag \\
&\;+ \frac{2}{dE_{i+1,j,k} + dE_{i,j,k}} \notag \\
&\;\quad \times \Bigg[ r_a (dE_{\mathrm{loss}})_{i+1,j,k} (N_e)_{i+1,j,k} \notag \\
&\phantom{\;\;\; \times \Bigg[}
    {} - \frac{1}{r_a} (dE_{\mathrm{loss}})_{i-1,j,k} (N_e)_{i-1,j,k} \Bigg] \notag \\
&\;+ (1 + z - \beta)(N_e)_{i,j-1,k} \notag \\
&\; + (\beta + \gamma - \eta)(N_e)_{i,j,k+1} \notag \\
&\;+ (\beta - \gamma + \eta)(N_e)_{i,j,k-1} \notag \\
&\; - 2 (\nabla \cdot \mathbf{V})_{i,j,k} \Delta t (N_e)_{i,j,k},
\end{align}
where $\beta$,  $\gamma$,  $\eta$,  and $z$ describe the evolution of particles with respect to time and energy on the discretized grid, and $r_a$ is introduced to regulate the energy-loss term.  $dE_{\rm loss}$ denotes the energy lost by particles during the time step $\Delta t$.  When used as subscripts,  $i$,  $j$, and $k$ represent the grid indices in energy, time, and space.  A detailed discussion of the derivation of Equation~(\ref{eq13}),  the implementation of the boundary conditions, the treatment of the energy-loss term, and the discretization methods in time,  energy,  and space can be found in Section~2 of  \citet{Xian2026ApJ}.

The evolution of the magnetic field is given by the following equation \citep{van2018arXiv}
\begin{equation}
\label{eq16}
B(r, t) = B_{\mathrm{age}} \left( \frac{r}{r_0} \right)^{\alpha_B} \left( \frac{t}{t_{\mathrm{age}}} \right)^{\beta_B},
\end{equation}
where $r_0$ is the fixed numerical inner boundary of the resolved transport region (see Figure~\ref{figure2}).  
$B_{\rm age}$ represents the effective magnetic-field strength adopted in the resolved large-scale transport region at the present age, rather than the magnetic field in the immediate vicinity of the termination shock. The magnetic field near the termination shock can be substantially higher, but it may be modified during the downstream evolution through processes such as shock compression, magnetic dissipation, and flow evolution.
$\alpha_B$ describes the spatial dependence of the magnetic field.  $\beta_B$ describes its temporal dependence.
Since no non-thermal radio or X-ray observations spatially matched to the extended component C1 are currently available, the radial magnetic-field distribution cannot be independently constrained. We therefore fix $\alpha_B = 0$, corresponding to no significant radial magnetic-field gradient in the present model, consistent with previous spatially dependent PWN modelling \citep{van2018MN}.  For the temporal evolution, previous PWN studies have commonly adopted $\beta_B$ values in the range $-1.15$ to $-1.45$, with $\beta_B = -1.3$ being a representative choice (\citealp{van2018MN}; \citealp{Kundu2024}).  We therefore fix $\beta_B = -1.3$ in the present work.
The radial dependence of $N_{\rm e}(r,E_{\rm e},t)$, $\kappa(r,E_{\rm e})$, and $B(r,t)$ therefore allows the model to account for the radial inhomogeneity of particle transport and local radiative properties within the resolved nebula.

\section{Modelling results}
\label{sect:3}

\subsection{Component C0 }
\label{sect:3.1}

\begin{figure}[t]
    \centering
    \includegraphics[width=0.95\columnwidth]{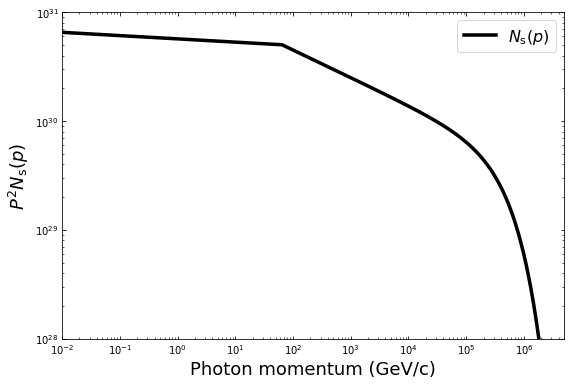}
    \caption{Momentum distribution of particles escaping from component C0  ($p_{\rm cut}=5.0\times10^{5}$~GeV/$c$).}
    \label{figure3}
\end{figure}

\begin{figure}[t]
    \centering
    \includegraphics[width=\columnwidth]{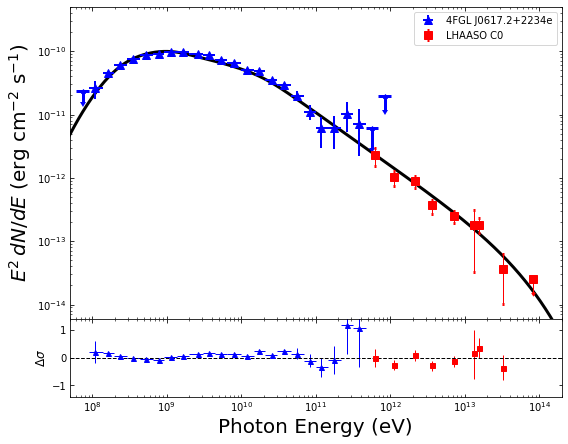}
    \caption{Upper panel: SED fit to component C0 . Lower panel: normalized residuals between the observed data and the best-fit model. The observational data are taken from \citet{LHAASOCollaboration2025}.}
    \label{figure4}
\end{figure}

In this section,  we apply the SNR--MC hadronic model introduced in Section~\ref{sect:2.1} to component C0  of IC~443.  To ensure consistency in geometric scale and physical scenario,  we adopt $d=1.5$~kpc and $t_{\rm age}=30$~kyr as the distance and age parameters for both components C0 and C1  in the parameter fitting presented in Section~\ref{sect:3.1} and Section~\ref{sect:3.2} (\citealp{Woltjer1972}; \citealp{Olbert2001}; \citealp{Prabir2024}).

The spatial extent of the external MC around IC~443 is taken to be $L_1=11.5$~pc and $L_2=14.7$~pc \citep{Ohira2011}. We adopt an effective target gas density of $n_{\rm H} = 20.0~{\rm cm}^{-3}$, following \citet{Alfaro2025}.  This value is the same as that adopted by \citet{Fermi2013}, who estimated it from a shock-disturbed gas mass of approximately $10^3~M_{\odot}$ in IC 443 and a characteristic remnant size of about 10 pc. Thus, the characteristic mass of the shocked molecular material associated with the adopted effective target density is of the order of $10^3~M_{\odot}$. In the present model, the molecular-cloud mass is not treated as an independent fitting parameter; instead, the target material for $pp$ interactions is characterised by the effective gas density $n_{\rm H}$, while the fraction of escaped particles interacting with the molecular cloud is described by $\eta$.
The relevant parameters adopted in the model calculations are listed in Table~\ref{tab1}.  Figure~\ref{figure3} shows the energy spectrum of particles escaping from the SNR at $t=t_{\rm age}$.  Using the Naima package in Python, the final SED of component C0  can be calculated from Equation~(\ref{eq3})  \citep{Zabalza2015}. As shown in Figure~\ref{figure4}, the SNR--MC hadronic model reproduces the observed flux of component C0  well, and the resulting SED exhibits a pronounced $\pi^0$-decay bump.

The fitting results yield a spectral index of $s=2.03$, which is close to the value obtained by \citet{Ohira2011}. In addition, the diffusion-coefficient normalization factor $\chi=0.08$ indicates that CR diffusion around component C0  is significantly suppressed, thereby increasing the rate of pp interactions between the particles and the surrounding MCs  \citep{Xian2025RAA}. As shown in Figure~\ref{figure3}, the energies of particles escaping from component C0  of IC~443 reach the PeV scale, suggesting that this component may be a potential PeVatron candidate.

\begin{table}
\caption{Adopted parameter values for component C1 }
\label{tab2}
\centering
\small
\setlength{\tabcolsep}{3pt}
\renewcommand{\arraystretch}{1.05}
\begin{tabular}{p{5cm}cc}
\hline\hline
Input Parameters                                  & Symbol                & Value \\
\hline
CMB Temperature (K)                               & $T_{\rm CMB}$         & 2.72 \\
CMB Energy Density  (eV cm$^{-3}$)                & $U_{\rm CMB}$         & 0.261 \\
FIR Temperature (K)                               & $T_{\rm FIR}$         & 20.0 \\
FIR Energy Density (eV cm$^{-3}$)                 & $U_{\rm FIR}$         & 0.75\\
NIR Temperature (K)                               & $T_{\rm NIR}$         & 3000 \\
NIR Energy Density (eV cm$^{-3}$)                 & $U_{\rm NIR}$         & 1.26 \\
Present-day spin-down luminosity (erg s$^{-1}$)   & $L_{\rm age}$         & $1.0 \times 10^{36}$ \\
Period (ms)                                       & $P$                   & 250.0 \\
Period derivative (s s$^{-1}$)                    & $\dot{P}$             & $1.27 \times 10^{-13}$ \\
Braking index                                     & $n$                   & 3 \\
Distance (kpc)                                    & $d$                   & 1.5 \\
Diffusion normalization in Region II              & $D_2$                 & $1.0 \times 10^{28}$ \\
\hline
\multicolumn{3}{l}{Fitted Parameters} \\
\hline
Present-day magnetic field ($\mu$G)              & $B_{\rm age}$          & 4.7 \\
Diffusion normalization in Region I              & $D_1$                  & $1.0 \times 10^{26}$ \\
Injection spectral index                         & $s$                    & 1.90 \\
Cutoff energy (eV)                               &$E_{\rm cut}$           & $1.75 \times 10^{12}$ \\
Particle energy conversion                       & $\epsilon$             & 0.6 \\
Diffusion index                                  & $\delta$               & 1.0 \\
\hline
\end{tabular}
\end{table}

\subsection{Component C1}
\label{sect:3.2}

In this section, we apply the multi-zone leptonic model introduced in Section~\ref{sect:2.2} to component C1  of IC~443 to describe the evolution of high-energy electrons in this region and the resulting $\gamma$-ray emission.  The radio-quiet neutron star CXOU~J061705.3+222127 is commonly regarded as a possible counterpart of this source. According to \citet{Din2026}, its spin period is $P=250$~ms, and its current spin-down luminosity $L_{\rm age}$ is approximately $(1$--$30)\times10^{36}$~erg~s$^{-1}$.  Using the relation between $P$ and $\dot{P}$, the period derivative is estimated to be $\dot{P}=1.27\times10^{-13}$~s~s$^{-1}$.  In this work, we adopt $L_{\rm age}=1\times10^{36}$~erg~s$^{-1}$. In the model calculations, we assume a dipole magnetic field and adopt a braking index of $n=3$ \citep{Gong2026}.

We use the particle spectrum evolved to the current age $t=t_{\rm age}$ as input to Naima to calculate the SED of component C1  \citep{Zabalza2015}.  In this model, the low-energy emission is mainly produced by synchrotron radiation, whereas the high-energy $\gamma$-ray emission is primarily generated through IC scattering of high-energy electrons off the background photon fields. The background photon fields considered in this work include the cosmic microwave background, the far-infrared photon field, and the near-infrared photon field, with corresponding temperatures and energy densities of $T_{\rm CMB}=2.72$~K and $U_{\rm CMB}=0.261$~eV~cm$^{-3}$, $T_{\rm FIR}=20$~K and $U_{\rm FIR}=0.75$~eV~cm$^{-3}$, and $T_{\rm NIR}=3000$~K and $U_{\rm NIR}=1.26$~eV~cm$^{-3}$ \citep{Porter2017}.

\begin{figure}[!t]
    \centering
    \includegraphics[width=\columnwidth]{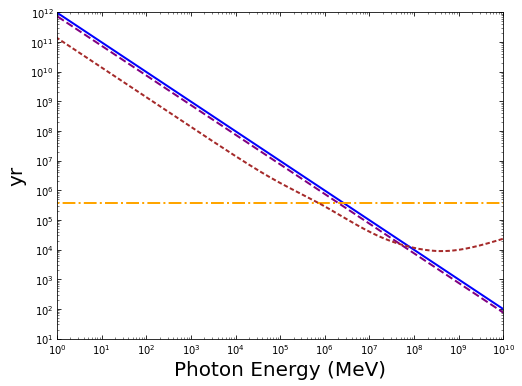}
    \caption{Characteristic timescales for particle transport and radiative cooling at $t = 30$ kyr and $r = 0.67^\circ$, including diffusion (blue solid line), advection (orange dash-dotted line), synchrotron radiation (purple dashed line), and IC  scattering (red dotted line).}
    \label{figure5}
\end{figure}

\begin{figure*}[t]
    \centering
    \includegraphics[width=0.49\textwidth]{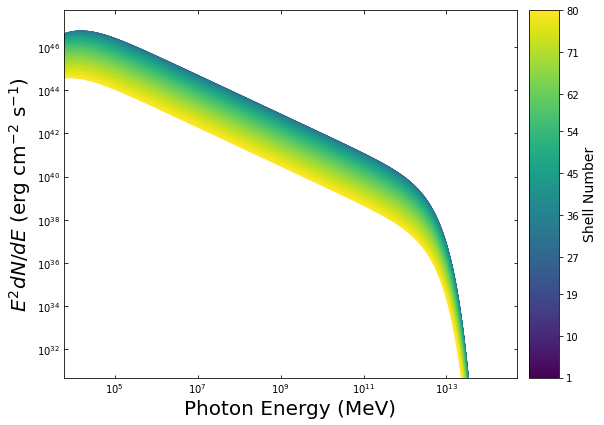}
    \hfill
    \includegraphics[width=0.49\textwidth]{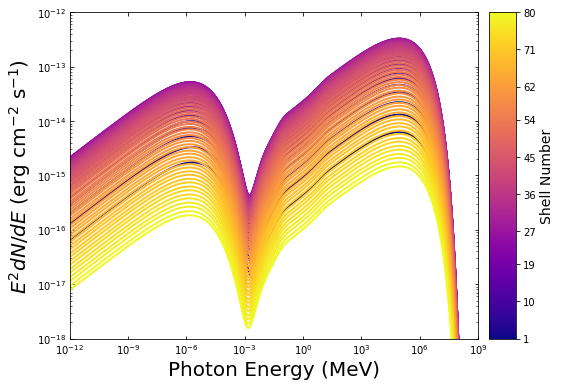}
    \caption{Left panel: electron spectra in successive radial shells from the centre outward.  Right panel: corresponding SEDs for the same shells in the same order.}
    \label{figure6}
\end{figure*}

These photon fields represent large-scale Galactic background radiation fields and are therefore approximated as spatially uniform and isotropic external target fields on the scale of the C1 region. This approximation applies only to the external background photon fields and does not imply that the electron distribution, magnetic field, or local emissivity inside the nebula is spatially uniform. In the multi-zone calculation, the electron distribution and radiative emissivity are evaluated shell by shell using the local values of $N_{\rm e}(r,E_{\rm e},t)$ and $B(r,t)$.

The model parameters adopted in the calculations are listed in Table~\ref{tab2}. Figure~\ref{figure5} shows the characteristic timescales associated with particle transport and radiative cooling at $t = t_{\rm age}$.  We further investigate the spatial evolution of the electron spectra along the radial direction, and the corresponding results are presented in  Figure~\ref{figure6}. The figure shows the electron spectra from shell~1 to shell~80 moving outward from the source centre, together with the corresponding multiwavelength SEDs. Both the electron spectra and the SEDs exhibit clear and systematic evolutionary trends with radius, indicating significant differences among the shells. Compared with the inner shells close to the source centre, the outer shells show a markedly lower overall normalized flux. This result indicates that the overall radiative flux and energy output of the system decrease progressively with increasing radius.

\begin{figure*}[t]
    \centering
    \includegraphics[width=0.49\textwidth]{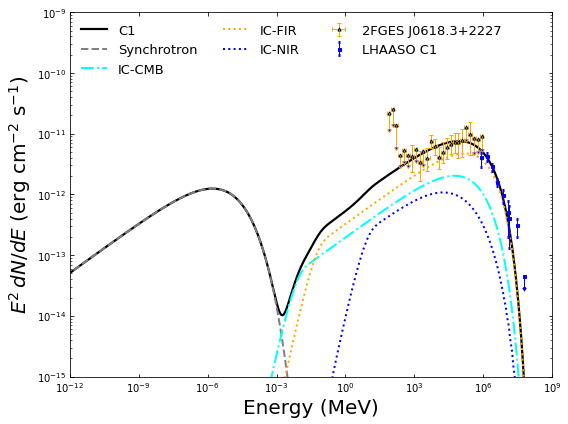}
    \hfill
    \includegraphics[width=0.49\textwidth]{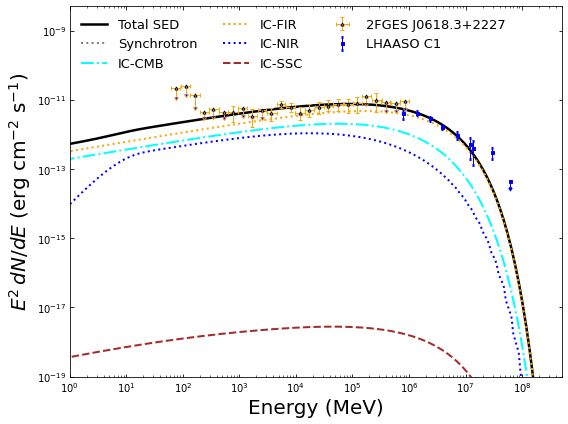}
    \caption{Left panel: overall multiwavelength SED of component C1 ($r \leq 0.67^\circ$).  Right panel: enlarged view of the $\gamma$-ray SED. The different radiative components correspond to synchrotron radiation (dark grey solid line), the cosmic microwave background (cyan dash-dotted line), the far-infrared photon field (orange dotted line), the near-infrared photon field (blue dotted line),  and the synchrotron self-Compton (brown dashed line).  The observational data are from \citet{LHAASOCollaboration2025}.}
    \label{figure7}
\end{figure*}

Using the parameters listed in Table~\ref{tab2}, we compute the multiwavelength SED of component C1  from Equation~(\ref{eq15}), and the corresponding results are shown in Figure~\ref{figure7}. The $\gamma$-ray flux predicted by the model is consistent with the Fermi-LAT and LHAASO observations of the C1 region. At the high-energy end, the SED exhibits a pronounced cooling cutoff, likely reflecting spectral steepening caused by long-term radiative losses of high-energy electrons. 
We note that the synchrotron component shown in Figure~\ref{figure7} is a model prediction rather than a component directly constrained by C1-specific low-energy observations. Although IC 443 has been extensively observed at radio and X-ray wavelengths, no spatially matched diffuse non-thermal radio/X-ray SED or stringent X-ray upper limit is currently available for the extended component C1. The detected non-thermal X-ray emission is associated with the much more compact PWN around CXOU J061705.3+222127, whereas the larger-scale X-ray emission is dominated by thermal emission from the SNR  \citep{Swartz2015}.

\begin{figure*}[t]
    \centering
    \includegraphics[width=0.49\textwidth]{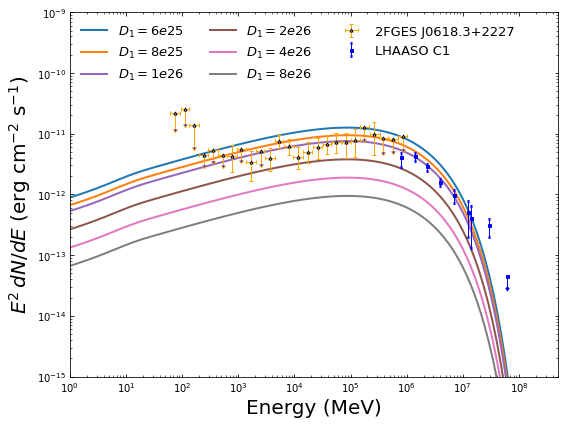}
    \hfill
    \includegraphics[width=0.49\textwidth]{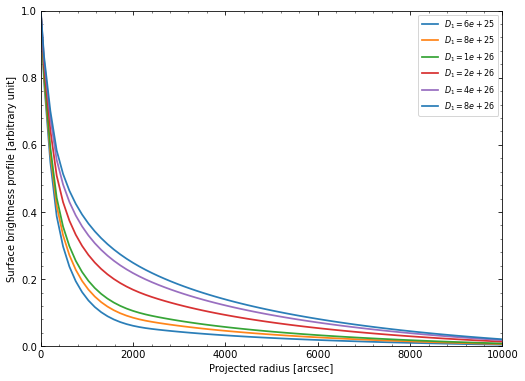}
    \caption{Dependence of component C1 ($r \leq 0.67^\circ$) SED and radial surface-brightness profile on the diffusion coefficient $D_1$.  Left panel: modelled SEDs for different values of $D_1$.  Right panel:  corresponding radial surface-brightness profiles of \(6\text{--}50\,\mathrm{TeV}\) emission. }
    \label{figure8}
\end{figure*}

In addition to the Galactic background photon fields, we have also assessed the possible contribution from the internal synchrotron photon field. This photon field gives rise to the synchrotron self-Compton (SSC) component. Internal radiation fields may be important in young and compact PWNe (\citealp{Amato2003}; \citealp{Bednarek2003}). However, component C1  is a large-scale extended source, and IC 443 is a middle-aged system. The energy density of the synchrotron photon field can be approximated as follows:
\begin{equation}
\label{eq17}
U_{\rm syn} \sim \frac{L_{\rm syn}}{4\pi R^2 c}
\end{equation}
where $L_{\rm syn}$ is the synchrotron luminosity, $R$ denotes the characteristic radius of the synchrotron-emitting region, and $c$ is the speed of light.  As the size of the emitting region increases, the energy density of synchrotron photons within the system decreases rapidly. In addition, no significant synchrotron counterpart associated with component C1  has been detected to date. The energy density of the internal synchrotron photon field in C1 is therefore expected to be much lower than that in young and compact PWNe.

To verify this expectation, we further calculated the SSC component using naima and the best-fitting electron distribution at $t = t_{\rm age}$. As shown in the right panel of Figure~\ref{figure7}, the SSC component is much weaker than the external IC emission produced by electrons scattering off the CMB, FIR, and NIR photon fields. This indicates that the internal synchrotron photon field neither dominates the radiative cooling of the C1 electrons nor makes the dominant contribution to the predicted $\gamma$-ray SED. In the present model, the SSC component is evaluated only as an a posteriori consistency check and is not included as an additional major cooling channel in the baseline transport calculation.

In IC 443, shock-heated molecular clouds or dust may produce significant infrared emission, although the corresponding radiation field is likely to be highly spatially inhomogeneous. For the C1 region, current observations do not provide sufficiently robust constraints on the corresponding energy density, $U_{\rm cloud}$. In its morphological analysis and spectral fitting,  \citet{LHAASOCollaboration2025} treated C0 and C1 as two independent $\gamma$-ray components. Based on this result, we further model C0 and C1 as two independent $\gamma$-ray sources and assume that their high-energy emission has different physical origins. For component C1 , we adopt a multi-zone leptonic model and assume that the radiating electron population is primarily associated with a PWN or TeV halo. Within this modelling framework, we do not further consider the coupling between the internal radiation fields and particle transport.  We neglect possible radiation produced by the SNR itself, as well as contributions from other potential radiation fields within the remnant.

The fitting results yield an injection spectral index of $s=1.90$ and a current magnetic field strength of $B_{\rm age}=4.7$~$\mu$G. These values are broadly consistent with those adopted by \cite{Liu2019} in their modelling of the typical TeV halo source Geminga, indicating that the physical quantities derived in this work are reasonable in magnitude. 
The left panel of Figure~\ref{figure8} shows the dependence of the modelled SED on the diffusion coefficient $D_1$. The best fit is obtained for $D_1=1\times10^{26}$~cm$^2$~s$^{-1}$,  which is approximately two orders of magnitude lower than the value commonly adopted for particle transport in the Galactic interstellar medium. This result indicates that particle diffusion is strongly suppressed in the C1 region. Such inefficient diffusion inhibits the rapid outward escape of high-energy electrons, allowing them to accumulate in the vicinity of the source and to sustain high-energy emission over an extended period. This confinement provides favourable conditions for the gradual formation of a TeV halo. Following the method proposed by \citet{van2018MN},  we calculated the radial surface-brightness profile as a function of distance from the source centre.  The right panel of Figure~\ref{figure7} compares the radial profiles obtained for different values of $D_1$,  illustrating the effect of the diffusion coefficient on the spatial extension of the emission region and on the radial decline in surface brightness.

\begin{figure}[t]
    \centering
    \includegraphics[width=\columnwidth]{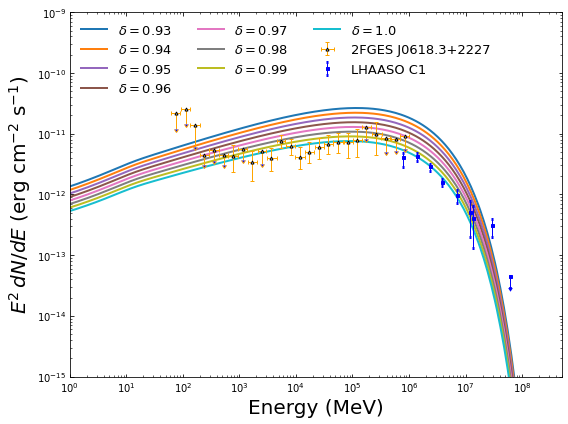}
    \caption{SED variation of component C1 ($r \leq 0.67^\circ$)  for different values of the diffusion index $\delta$.}
    \label{figure9}
\end{figure}

\begin{figure}[t]
    \centering
    \includegraphics[width=\columnwidth]{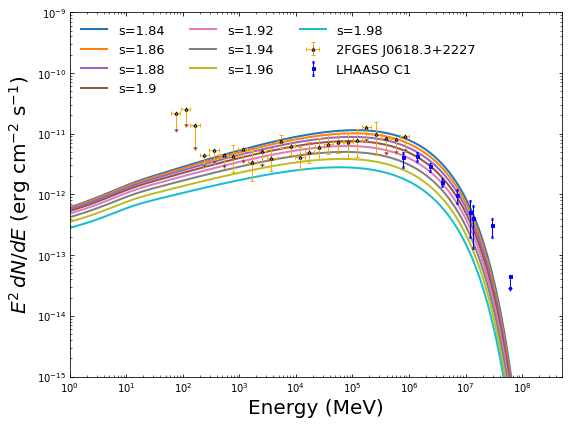}
    \caption{SED variation of component C1 ($r \leq 0.67^\circ$)  for different values of the injection spectral index $s$.}
    \label{figure10}
\end{figure}

To assess the influence of different model parameters on the SED, we further explore the ranges of $\epsilon$, $\delta$, and $s$. Over the parameter range considered here, the SED of C1 depends only weakly on $\epsilon$. Varying $\epsilon$ mainly produces a modest change in the overall normalisation of the radiative flux without significantly affecting the spectral shape.  Figure~\ref{figure9} presents the SEDs for different values of $\delta$. The model best reproduces the observations for $\delta=1$, indicating that particle diffusion in the C1 region is closer to a Bohm-like diffusion regime. Figure~\ref{figure10} shows the SED variations for different values of $s$ with all other parameters held fixed. As $s$ increases, the relative contribution of high-energy particles decreases, in agreement with the physical expectation for an injected particle population that follows a power-law distribution.

\section{Summary and Discussion}
\label{sect:4}

Based on the decomposition of the high-energy $\gamma$-ray emission in the IC 443 region reported by \citet{LHAASOCollaboration2025}, we model the C0 and C1 components independently because they differ markedly in spatial morphology and spectral properties.  For component C0,  which is located inside the remnant and has a relatively compact spatial distribution, we adopt an SNR-MC hadronic model that takes finite source-size effects into account, and suggest that it may be a PeVatron candidate.  For the extended component C1,  we employ a spatially dependent multi-zone leptonic model and further suggest that it may be a  TeV halo candidate.  The two models reproduce the respective SEDs of C0 and C1 reasonably well.

From a theoretical perspective, our results indicate that the origin of the $\gamma$-ray emission from the classical mixed-morphology middle-aged supernova remnant IC 443 is unlikely to be fully described by a single radiation mechanism, but instead requires different physical scenarios to be considered for its two $\gamma$-ray components.  This suggests that IC~443 is not only an important object for studying SNR-MC interactions, but may also serve as a representative case linking three classes of high-energy phenomena, namely SNRs,  PWNe, and TeV halos. These results provide a viable theoretical avenue for understanding the formation and evolution of $\gamma$-ray emission mechanisms in complex physical environments.

Although components C0 and C1  are described using hadronic and leptonic models, respectively, this does not imply that only one particle population is present in either region. For component C0, in addition to the high-energy protons responsible for the dominant $\pi^{0}$-decay $\gamma$-ray emission, primary electrons accelerated at the SNR shock and secondary electrons and positrons produced in $pp$ interactions may also contribute to the non-thermal emission. In particular, previous calculations have shown that, for sufficiently strong post-shock magnetic fields, synchrotron radiation from secondary electrons and positrons can make an important contribution to the observed radio emission of IC 443 \citep{Huang2020}. Since the $\gamma$-ray SED of component C0 is already well reproduced by the SNR-MC hadronic model, we do not introduce an additional independent leptonic $\gamma$-ray component in the present calculation.

A hadronic contribution to C1 cannot be completely excluded either.  \citet{LHAASOCollaboration2025} showed that the broadband $\gamma$-ray SED of C1 can also be described in a hadronic scenario and suggested that the difference between the proton spectra of C0 and C1 may be related to particle propagation. However, the $\gamma$-ray morphology of C1 does not show a clear spatial correlation with the molecular-gas distribution, while a leptonic IC scenario can also reproduce its observed properties. We therefore adopt the leptonic model as the baseline interpretation of C1, while allowing for a possible hadronic contribution. The available data are currently insufficient to constrain independently the composition of heavier of nuclei in C1, and such a component is therefore not included explicitly. A complete hybrid treatment of these additional particle populations would require further constraints on the particle composition, gas distribution, and multiwavelength emission and is beyond the scope of the present work.

Nevertheless, the SNR-MC hadronic model and the multi-zone leptonic model adopted in this work still have some limitations.

The SNR-MC hadronic model is built on the idealized assumptions of spherical symmetry and a homogeneous medium. However, in the real Galactic environment, the magnetic-field structure usually exhibits a highly complex three-dimensional configuration, including both large-scale ordered-field components and multiscale turbulent structures. Meanwhile, MCs themselves generally show pronounced clumpiness and density inhomogeneity, and the distribution of dense material and magnetic-field fluctuations within them can significantly affect CR diffusion and the regions where pp interactions occur. In addition, the morphology of an SNR, the direction of shock propagation, and its relative position with respect to the surrounding clouds are often not strictly symmetric, and in some cases significant anisotropic diffusion and escape channels may also exist \citep{Ustamujic2021}. Therefore, the present simplifications in geometry and medium conditions still impose certain limitations on the model, and this also calls for the further development of more refined three-dimensional numerical models.

On the other hand, the multi-zone leptonic model assumes that the observed radiation mainly originates from the associated pulsar, thus neglecting any additional contribution that may arise from the surrounding SNR. In addition, the spherical symmetry adopted in this model does not take into account the proper motion of the pulsar itself or the inhomogeneity of the surrounding environment \citep{Xian2026ApJ}.
In particular, while the present model includes radial inhomogeneity in particle transport, magnetic-field structure, and local emissivity, it does not self-consistently follow the full three-dimensional structure and time evolution of internal radiation fields.
If future studies can combine higher angular-resolution $\gamma$-ray observations, deeper X-ray and radio data, and more refined three-dimensional particle-transport simulations to place joint constraints on the spatial and spectral properties of C0 and C1, it will be possible to further clarify the relative contributions of the hadronic and leptonic components in the IC~443 region and to deepen our understanding of the near-source transport and radiative processes of high-energy particles in the Galaxy.

\begin{acknowledgements}
We sincerely thank the anonymous referee for the constructive comments.
JF is supported by the National Natural Science Foundation of China (grant Nos. 12563010 and 12393852), the Yunnan Fundamental Research Projects (grant No. 202501AS070068), and the International Centre of Supernovae (ICESUN) of the Yunnan Key Laboratory of Supernova Research (grant No. 202505AV340004). BW is supported by the National Natural Science Foundation of China (No 12225304), the Yunnan Revitalization Talent Support Program (Yunling Scholar Project), the Yunnan Science and Technology Program (Nos  202605AS350010 and 202601BC070011).  Y.J.Xian is supported by the Graduate Research Innovation Fund of Yunnan University (grant No. KC-262615769).
\end{acknowledgements}

\bibliographystyle{aa}
\bibliography{ams}

\end{CJK*}
\end{document}